\pdfoutput=1

\documentclass[11pt]{article}

\usepackage[preprint]{acl}
\usepackage{times}
\usepackage{algorithm}
\usepackage{algpseudocode}
\usepackage{latexsym}
\usepackage{multirow}
\usepackage{array}
\usepackage{subcaption}
\usepackage{tabularx}
\usepackage{svg}
\usepackage{hyperref}
\usepackage{mathtools}
\usepackage{graphicx}
\usepackage{caption}
\usepackage[labelformat=simple]{subcaption}

\usepackage[T1]{fontenc}

\usepackage[utf8]{inputenc}

\usepackage{microtype}

\usepackage{inconsolata}

\definecolor{midnightgreen}{rgb}{0.0, 0.29, 0.33}

\title{Enhancing Dense Retrievers' Robustness with Group-level Reweighting}

\author{Peixuan Han\thanks{Part of this work was done while the author was visiting Carnegie Mellon University.} \\
  Tsinghua University \\
  \texttt{hanpx20@mails.tsinghua.edu.cn} \\\And
  Zhenghao Liu \\
  Northeastern University \\
  \texttt{liuzhenghao@mail.neu.edu.cn} \AND
  Zhiyuan Liu \\
  Tsinghua University \\
  \texttt{liuzy@tsinghua.edu.cn} \\\And
  Chenyan Xiong \\
  Carnegie Mellon University \\
  \texttt{cx@cs.cmu.edu}}

\begin{document}
\maketitle

\begin{abstract}
The anchor-document data derived from web graphs offers a wealth of paired information for training dense retrieval models in an unsupervised manner. However, unsupervised data contains diverse patterns across the web graph and often exhibits significant imbalance, leading to suboptimal performance in underrepresented or difficult groups. In this paper, we introduce \texttt{WebDRO}, an efficient approach for clustering the web graph data and optimizing group weights to enhance the robustness of dense retrieval models. Initially, we build an \textbf{embedding model} for clustering anchor-document pairs. Specifically, we contrastively train the embedding model for link prediction, which guides the embedding model in capturing the document features behind the web graph links. Subsequently, we employ the group distributional robust optimization to recalibrate the weights across different clusters of anchor-document pairs during training \textbf{retrieval models}. During training, we direct the model to assign higher weights to clusters with higher loss and focus more on worst-case scenarios. This approach ensures that the model has strong generalization ability on all data patterns. Our experiments on MS MARCO and BEIR demonstrate that our method can effectively improve retrieval performance in unsupervised training and finetuning settings. Further analysis confirms the stability and validity of group weights learned by \texttt{WebDRO}. All codes will be released via GitHub.
\end{abstract}

\section{Introduction}
Dense retrieval models, which encode the semantics of text into dense vectors, have become prevalent and achieved state-of-the-art performance on many public benchmarks thanks to the surge of pre-trained language models like BERT~\cite{devlin2018bert} and T5~\cite{raffel2020exploring}. With enough annotated data, dense retrieval models can adeptly capture semantic features in supervised scenarios, outperforming traditional sparse methods such as BM25~\cite{robertson2009probabilistic, robertson1995okapi}, which solely retrieves documents containing keywords already present in the query. However, annotated data is inaccessible in many scenarios either due to the extensive effort needed for annotation or privacy concerns. This highlights the necessity for unsupervised dense retrievers. Several paradigms of unsupervised dense retrieval have been proposed, among which leveraging \textbf{web anchors} is a promising choice. Primarily, anchor-document pairs contain weak supervision signal that indicates overlapping information~\cite{xie2023doremi, zhang2020selective} and thus can approximate real search queries better than plain-text-based methods like ICT~\cite{izacard2021unsupervised}. In addition, anchor resources are abundant across the Internet, bearing greater potential for the scalability of retrievers compared to methods based on the title and abstract of the passage.

Although the web graph contains rich data patterns, the imbalanced data distribution is an inherent flaw of unsupervised retrieval data. Training robust dense retrievers with these data is challenging, as different patterns exhibit significantly different frequencies in training data, leading to imbalanced representation. Due to insufficient exposure to underrepresented data patterns, the model's performance on downstream tasks involving similar patterns is hindered. To mitigate this, previous research has primarily focused on instance-level data selection to identify data points whose patterns haven't been sufficiently trained by the model. To reach a more tailored selection, these approaches typically leverage feedback signals during the training process to assign a weight to each instance, such as Active Learning~\cite{settles2011theories,coleman2019selection} and Meta Adaptive Ranking~\cite{sun2020few}. These methods require the model to be trained multiple times, which is highly time-consuming and computationally intensive.


We address the lack of both efficient and effective retrieval data reweighting methods by introducing a \textbf{group-level clustering and reweighting strategy}--\texttt{WebDRO}. \texttt{WebDRO} enables us to conduct data reweighting in a single pass, significantly saving computational resources. Additionally, data clustered into groups encapsulates much more information than a single data point and represents the common characteristics of the same pattern, \texttt{WebDRO} can enhance the quality and stability of reweighting.

Specifically, we employ a two-step approach to train an unsupervised dense retrieval model using \texttt{WebDRO}. The first step (which we denote as \texttt{Step A}) is to generate clusters from unsupervised web data. We utilize structural information from the web graph to train an \textbf{embedding model} with contrastive learning (\texttt{Step A-1}). By distinguishing related web pages from unrelated ones, the embedding model can work as a link predictor that captures the semantic features of web documents. Leveraging the embedding model, we generate vector representations for documents in the anchor-document pairs and produce high-quality clusters through K-Means (\texttt{Step A-2}). In the second step (which we denote as \texttt{Step B}), we utilize Group Distributionally Robust Optimization (\textbf{GroupDRO})~\cite{oren2019distributionally,sagawa2019distributionally} to reweight the clusters during the contrastive training of the \textbf{retrieval model}. Originally designed to tackle distributional shifts in language model training, GroupDRO dynamically adjusts the weights of different groups based on the training loss. It increases the weight of the groups with large training loss, therefore highlighting the underrepresented groups that the model has difficulty learning from and enhancing the model's robustness in versatile retrieval tasks.

We conduct experiments on two widely used retrieval datasets, MS MARCO~\cite{nguyen2016ms} and BEIR~\cite{thakur2021beir}. Results indicate that our method notably enhances retrievers' performance in both unsupervised and finetuned settings. Specifically, \texttt{WebDRO} yields a performance improvement of 1.2\% (measured by nDCG@10) on MS MARCO and 1.1\% on BEIR compared to our main baseline, Anchor-DR~\cite{xie2023unsupervised}. Analytical experiments further demonstrate that our model consistently produces stable, high-quality clusters and accurately identifies groups that baseline dense retrievers struggle with.

The main contributions of our work are:
\begin{itemize}
    \item We explored web anchors' usage in training dense retrievers and proposed \texttt{WebDRO}, a clustering and group-level reweighting algorithm for unsupervised dense retriever training.
    \item Empirical analysis shows that \texttt{WebDRO} can efficiently enhance dense retrievers' robustness, improving their performances on both unsupervised and finetuned settings.
    \item We further confirmed \texttt{WebDRO}'s ability to efficiently identify underrepresented and challenging groups, thereby helping the model improve its capabilities more effectively.
\end{itemize}

\section{Related Work}
This section describes relevant work, focusing on dense retrieval models and distributionally robust optimization.

\textbf{Dense Retrieval.} Dense Retrieval is an approach where both the query and document are separately encoded through a neural network. The prevalent training approach, introduced by \citet{karpukhin2020dense}, trains the network with a contrastive loss. 
Although dense retrieval has outperformed sparse methods~\cite{robertson2009probabilistic} in many tasks, its benefits are generally confined to well-annotated data. Unsupervised dense retrieval efforts focus on using unannotated data to generate query-document pairs that semantically resemble contrastive pairs in supervised scenarios. For instance, ICT~\cite{izacard2021unsupervised} samples a text span in a passage as the query and the rest part as the document, and Co-doc\cite{gao2021unsupervised} independently sample two text spans as the query and document. More recently, researchers have exploited unsupervised data at a greater scale and with more diverse constructing methods, namely E5~\cite{wang2022text} and BGE M3~\cite{chen2024bge}. In our work, we focus on one specific method of constructing contrastive pairs. Following Anchor-DR~\cite{xie2023unsupervised}, we use anchors and corresponding webpages in the Internet to train dense retrievers. Web anchors are proved to be a high-quality data source for unsupervised retrievers, and has great potential of scaling up, considering the ubiquity and large number of anchors available in the Internet.



\begin{figure*}
  \begin{subfigure}{\columnwidth}
    \includegraphics[width=\linewidth]{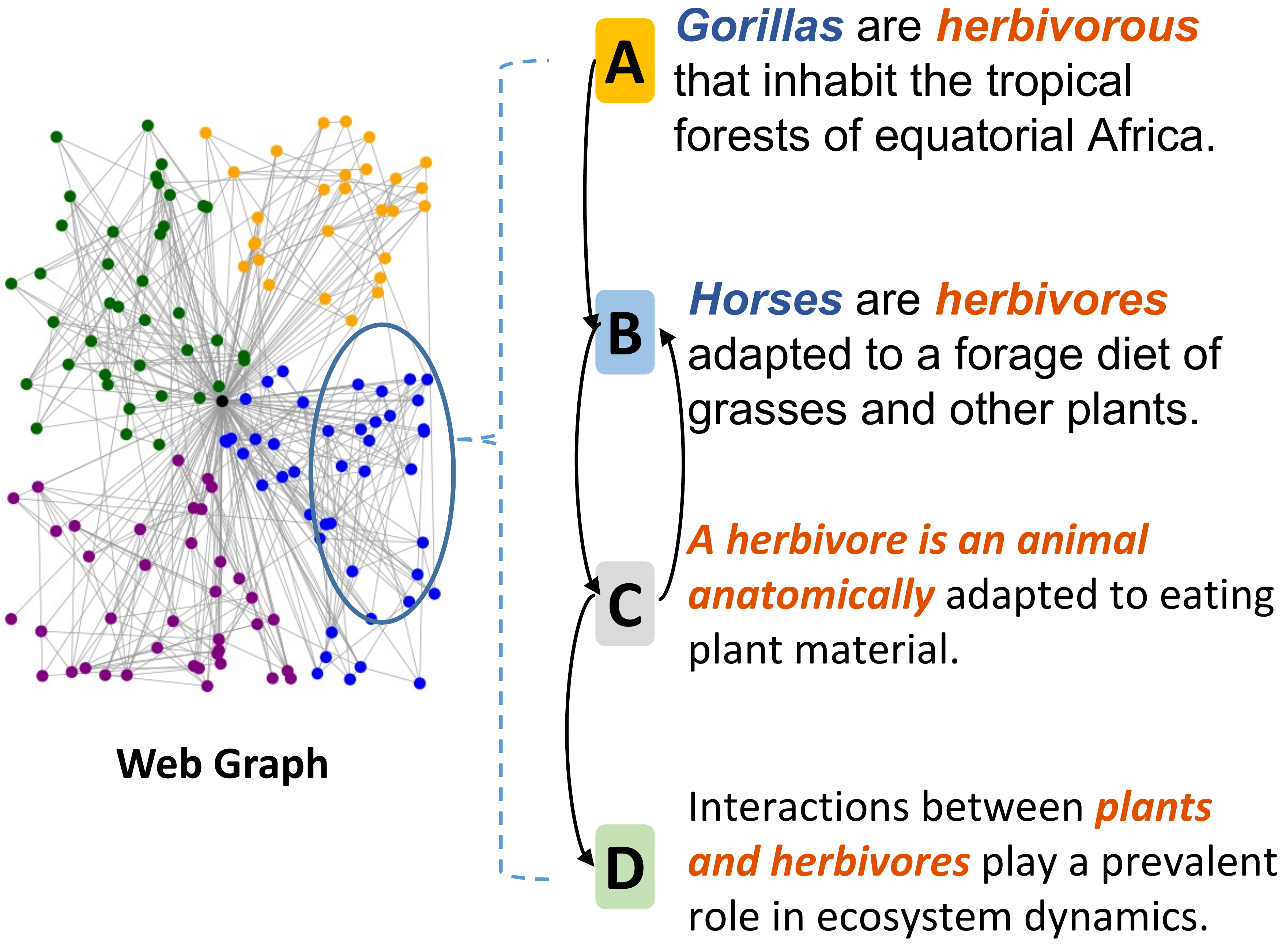}
    \caption{Web-based clustering.\label{fig:model:a}}
  \end{subfigure}
  \begin{minipage}{0.08\columnwidth}
  \hfill
  \end{minipage}
  \begin{subfigure}{0.9\columnwidth}
    \includegraphics[width=\linewidth]{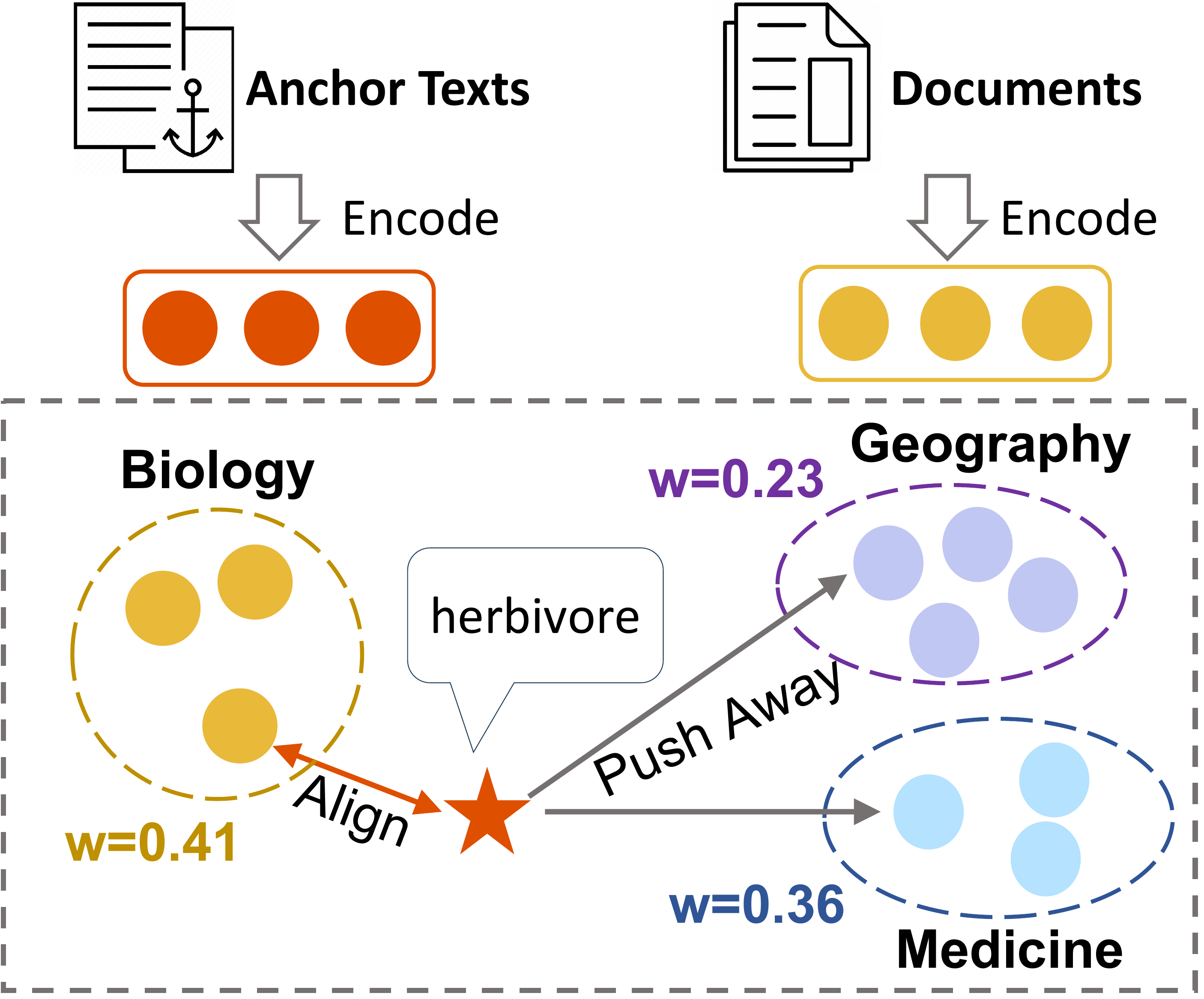}
    \caption{Contrastive training with \texttt{WebDRO}.  \label{fig:model:b}}
  \end{subfigure}
  \caption{\label{fig:model}Illustration of key steps in \texttt{WebDRO}. Figure~\ref{fig:model:a} shows the clustering process in \texttt{Step A}, where we use web links to train an embedding model and use the model to cluster documents. Figure~\ref{fig:model:b} shows the training process of the retrieval model in \texttt{Step B}. We aim to align the anchor with the corresponding document while pushing apart unrelated ones. Each group is assigned a weight that is dynamically updated based on the training loss.}
\end{figure*}

\textbf{Distributionally Robust Optimization.}
To address the issue of distribution shift~\cite{agarwal2021theory,wiles2021fine}, researchers have introduced Distributionally Robust Optimization (DRO)
~\cite{ben2013robust,rahimian2019distributionally,lin2022distributionally}. The goal of DRO is to optimize the worst-case performance of a model across various data distributions, ensuring its robustness against uncertain distributional changes. Specifically, GroupDRO considers robustness over groups and calculates the worst-case loss over groups by optimizing group weights and model weights concurrently. GroupDRO can be used to train more robust models, as well as to learn the importance of different data groups~\cite{xie2023doremi}. Prior applications of GroupDRO used explicitly pre-partitioned groups, and the number of groups is relatively small ($<50$). In this work, we extend the applicable scope of GroupDRO to a large number of groups clustered by implicit information.

\section{Methodology}
In this section, we first describe the preliminary of dense retrieval (Sec.~\ref{model:dr}).
Then we introduce our \texttt{WebDRO} method, which is shown in Figure~\ref{fig:model}.
In \texttt{Step A}, we use the method in Sec.~\ref{model:cluster} to train an \textbf{embedding model} using the web links. Then, we cluster the anchor-document pairs based on vector representations produced by the embedding model. In \texttt{Step B}, we train a robust dense \textbf{retrieval model} with Group Distributionally Robust optimization (GroupDRO) as introduced in Sec.~\ref{model:GroupDRO}.

\subsection{Preliminary of Dense Retrieval}\label{model:dr}
Given a query $q$, the dense retrieval task aims to search relevant document $p$ to satisfy the information needs of users. 

In dense retrieval, we train one single model to encode \textbf{both} queries and documents as dense vectors and then conduct relevance modeling in an embedding space. We use cosine similarity to measure the relevance between queries and documents. Through contrastive learning~\cite{oord2018representation,chen2020simple,he2020momentum}, the encoder is optimized to align the queries with related documents and pull away the irrelevant documents from queries. Specifically, the following loss function is minimized during training:
\begin{equation}\label{eq:contrast}
    \resizebox{0.8\linewidth}{!}{$
    l_{\text{dr}} (q)=-\log\frac{e^{sim(E(q), E(p^+))}}{e^{sim(E(q), E(p^+))} + \sum_{i=1}^k{e^{sim(E(q), E(p^-_i))}}},
    $}
\end{equation}

where $\tau$ is the temperature hyperparameter. $p^+$ stands for the related document for query and $p^-_1, ..., p^-_k$ are negative (unrelated) documents.

\subsection{\texttt{Step A}: Clustering with the Embedding Model}\label{model:cluster}
This subsection introduces our web graph based anchor-document cluster method. It learns the document representations using the document links in the web graph and then clusters the documents according to their semantic features.

\textbf{Document Relations in Web Graph.} The web graph models the document relationship with hyperlinks, which allows users to navigate from one web page to another or within the same web page. It is evident that the documents linked with a hyperlink will likely share similar or related information. As shown in Figure~\ref{fig:model:a}, these four documents are related and are linked with the anchor hyperlink, \textit{e.g.} the link between document A and document B indicate that ``horses and gorillas have some similarities'', showing an analogy relationship; the links between document B and document C indicate the definitions of ``horses'' and ``herbivorous'', showing the hypernymy relationship.

\textbf{Embedding Model Training.} We train an \textbf{embedding model} using the document relations in the web graph. We encode the documents using a pretrained language model encoder and contrastively train the encoder. A web page can contain multiple links, so we enumerate each link to construct contrastive pairs as training examples. For each link, we regard the source web page as the query, the target web page as the positive document, and sample unlinked documents as negative documents. The objective function is identical to that in Sec.~\ref{model:dr}. As a result, the \textbf{embedding model} is trained as a link predictor that can distinguish linked web pages from non-linked ones.

\textbf{Document Clustering.} Since the embedding model is capable of capturing webpages' features, we use it to encode all documents as dense vectors. Then we use Mini-Batch K-Means~\cite{sculley2010web}, a variant of K-Means~\cite{krishna1999genetic} to cluster embedding vectors of documents into a fixed number of groups. Some documents have more than one anchor that links to it, these anchor-document pairs are naturally clustered into the same group. To make sure each group has statistical significance, we set a threshold of the number of documents in a group as $MinSize=128$. All the groups with sizes smaller than $MinSize$ will be aggregated together and considered as a larger group. As this large group contains heterogeneous elements with different topics, we should not assign a single weight to represent this group. Therefore, the large group will not be considered in the reweighting discussed in Sec.~\ref{model:GroupDRO}.

\subsection{\texttt{Step B}: Robust Dense Retriever Training}\label{model:GroupDRO}
For each anchor in our corpus, we regard the text of the anchor as the query, and the content of the linked webpage as positive document. The anchor-document pairs contain different types of documents, and the distribution of different types could vary greatly: some occur frequently, while others are rare; some are easy for the model to learn, while others are challenging. Treating all data equally may affect the performance of dense retrievers when faced with underrepresented data types. To alleviate this problem, we follow~\citet{sagawa2019distributionally} to reweight different anchor-document groups and train more robust dense retrievers. 

Following GroupDRO, each group is assigned with a weight that indicates its importance, and the weights are dynamically updated during the training process. Specifically, we denote $n$ as the number of groups after clustering. Then, for a training instance $p$ in step $t$ that belongs to group $k$, the training loss of \texttt{WebDRO} is calculated as:
\begin{equation}\label{eq:groupdro_loss}
    L_\text{WebDRO}(p)=l_\text{dr}(p)\cdot w_k^t\cdot n \cdot C_k,
\end{equation}
where $l_\text{dr}$ is the contrastive loss calculated using Eq.~\ref{eq:contrast}, $w_k^t$ is the weight of the $k$-th group in step $t$, and $C_k$ is size factor of the $k$-th group. We will next describe $C_k$ and $w_k$ respectively. 

\textbf{Size Factor ($C_k$).} In a standard GroupDRO scenario, data are sampled from each group according to the uniform distribution. However, the size of different anchor-document groups in our method may vary significantly. To accommodate this circumstance, \texttt{WebDRO} introduces a size factor $C_k$:
\begin{equation}
    C_k=\frac{1}{N_k} \cdot \frac{\sum_{j=1}^n{N_j}}{n},
\end{equation}
where $N_k$ is the size of the $k$-th anchor-document group. Smaller groups will have a larger $C_k$ value, which serves as a balance for different group sizes. This is a natural extension of the original DRO, as $C_k=1$ for all $k$s when all groups have the same size (all $N_k$s are the same).

\textbf{Group Weighting ($w_k$).} The initial weight of all groups is the same: $w_k^0=\frac{1}{n}$. In each step, the groups' weights are updated \textbf{before} the calculation of $L_\text{WebDRO}$ (Eq.~\ref{eq:groupdro_loss}). First, we increase group weights based on training loss and get intermediate weights $w_k'$. For each group $k$, 
\begin{equation}
    w_k'=w_k^{t-1}\cdot e^{lr\cdot  C_k \cdot L_{\text{avg}k}}.
\end{equation}
In this formula, $lr$ is the learning rate of weight updating. $ L_{\text{avg}k}$ portraits the contribution of group $k$ in step $t$'s training loss, which can be calculated:
\begin{equation}
L_{\text{avg}k} = \frac{\sum_{p\text{ in group }k}{l_{\text{dr}}(p)}}{N^t},
\end{equation}
where $N^t$ is the number of training data in step $t$. This is a natural extension of the case when all training data in a step belongs to a single group. $L_{\text{avg}k}$ for that group will thus be the average loss, and other group weights will remain unmodified.

After each step, the sum of group weights should be $1$, so we normalize the intermediate weights to get $w_k^t$. For each group $k$,
\begin{equation}
    w_k^{t}=\frac{w_k'}{\sum_{j=1}^n{w_j'}}.
\end{equation}

During training, \texttt{WebDRO} aims to increase the weights of groups with higher contrastive loss, which leads the model to draw more attention to groups that are challenging for it. This enhances the model's robustness on underrepresented groups in the source, thereby implicitly improving the model's generalization ability on unseen target queries. Algorithm~\ref{alg:1} shows the process of weight updating, which is the same as the above description.

\begin{algorithm}
\caption{Weight Updating in \texttt{WebDRO}}
\label{alg:1}
\begin{algorithmic}[1]
    \State \textbf{Input:} Model $M$, step number $t$, number of groups $n$, learning rate for weight updating $lr$, size factor $C[1,\dots,n]$, size of batch $N^t$, data batch $p[1,\dots,N^t]$, original weights $w^{t-1}[1,\dots,n]$
    
    \State \textbf{Output:} Updated weights $w^{t}[1,\dots,n]$
    \\
    \State $L_{\text{avg}}[1, \dots, n] \gets [0, \dots, 0]$
    \For{$i = 1$ to $N^t$}
        \State Find the group number $k$ of $p[i]$
        \State Calculate the contrastive loss $l_{dr}(M, p[i])$
        \State $L_{\text{avg}}[k] \gets L_{\text{avg}}[k] + l_{dr}(M, p[i])$
    \EndFor
    \For{$k = 1$ to $n$}
        \State $L_{\text{avg}}[k] \gets L_{\text{avg}}[k]\ /\ N^t$
        \State $w'[k] \gets w^{t-1}[k] \cdot \exp(lr\cdot C[k]\cdot L_{\text{avg}}[k])$
    \EndFor
    \State $w_{\text{sum}} \gets \sum_{k=1}^n{w'[k]}$
    \For{$k = 1$ to $n$}
        \State $w^{t}[k] \gets w'[k]\ /\ w_{\text{sum}}$
    \EndFor
   \State \textbf{RETURN} $w^{t}[1,\dots,n]$
\end{algorithmic}
\end{algorithm}

\begin{table*}
    \setlength{\arrayrulewidth}{0.1mm}
    \centering
    \resizebox{\linewidth}{!}{%
    \begin{tabular}{l|c|cccccc|ccc}
    \hline
    \multirow{2}{*}{\textbf{Model}} & \multirow{2}{*}{\textbf{BM25}} & \multicolumn{6}{c|}{\textbf{Unsupervised Dense Retrievers}} & \multicolumn{3}{c}{\textbf{Finetuned Dense Retrievers}}\\
    & & SPAR & AugTriever & coCondenser & E5-Large & Anchor-DR & \textbf{WebDRO} & T5-Base & Anchor-DR & \textbf{WebDRO}
    \\\hline
        \# Training Pairs & - & 22.6M & 52.4M & 8.8M & 270M & 13.8M & 13.8M & - & - & -
    \\\hline
        \textbf{MS MARCO} &22.8&19.2&20.6&16.2&26.2&26.15 &\underline{27.37}&36.47 &39.13 & \textbf{40.58}\\
        trec-covid &65.6&53.1&53.5&40.4&61.8& \underline{72.16} &69.26 &72.43&73.98&\textbf{78.04} \\
        nfcorpus &\textbf{32.5}&26.4&30.3&28.9&\underline{33.7}& 30.70 &30.74 &29.22&29.75&31.24 \\
        fiqa &23.6&18.5&22.3&25.1&\underline{43.2}& 23.79 & 25.12&26.72&28.11&\textbf{28.45}\\
        arguana &31.5&42.0&39.1&\underline{44.4}&\underline{44.4}& 28.39 & 31.27&47.59&\textbf{49.18}&48.02 \\
        touche2020 &\textbf{36.7}&\underline{26.1}&21.6&11.7&19.8& 21.85 & 24.05&23.16&24.06&27.56  \\
        quora &78.9&70.4&82.7&82.1&\textbf{\underline{86.1}}& 85.57 & 83.36&85.28&85.48&85.85  \\
        scidocs &15.8&13.4&14.7&13.6&\textbf{\underline{21.8}}&13.42&14.93 &14.95 &15.41&15.30  \\
        scifact &66.5&62.6&64.4&56.1&\textbf{\underline{72.3}}& 58.84 & 61.88 &57.54 &61.06&62.16 \\
        nq &32.9&26.2&27.2&17.8&\underline{41.7}& 28.83 & 34.48&42.48&45.24&\textbf{47.23} \\
        hotpotqa & \textbf{60.3} & \underline{57.2} & 47.9 & 34.0 & 52.2 & 53.81 &52.27&50.25&56.90&57.39 \\
        dbpedia-entity &31.3&28.1&29.0&21.5&\underline{37.1}& 34.61 &35.34 &34.01&36.39&\textbf{38.09} \\
        fever &\textbf{75.3}&56.9&59.7&61.5&68.6& 72.15 &\underline{73.42} &68.42&71.69&70.93\\
        climate-fever &\textbf{21.4}&16.4&17.7&16.9&15.7& 18.87 & \underline{21.05} &18.45&18.95&18.89 \\
        cqadupstack &29.9&27.9&27.1&30.9&\textbf{\underline{38.9}}& 28.82 &32.97 &33.64&35.79&35.25\\
        trec-news &39.8&-&-&25.4&-& \underline{41.18} & 40.24&39.68&\textbf{43.05}&41.80\\
        signal1m &\textbf{33.0}&-&-&21.4&-& 22.99 &24.36 &26.48&27.56&27.82 \\
        bioasq &\textbf{46.5}&-&-&22.7&-& 33.88 & \underline{38.45}&30.07&33.73&35.86 \\
        robust04 &40.8&-&-&29.8&-& \underline{41.01}& 37.60&\textbf{44.19}&43.23&43.51 \\
    \hline
        \textbf{Avg. on BEIR14\footnotemark[1]} &43.0&36.2&37.0&33.4&\underline{44.2}& 40.84 & 42.15&43.15&45.14&\textbf{46.03}\\
        \textbf{Avg. All} &41.3&-&-&31.6&-& 38.80 & \underline{39.91}&41.11&43.09&\textbf{43.89} \\
    \hline
    \end{tabular}}
    \caption{Retrieval results of different models on MS MARCO and BEIR. All models are evaluated by nDCG@10. Model names in the "Finetuned Dense Retrievers" category only indicate the base model before finetuning. The best result on each task is marked in bold, and the best result among unsupervised dense retrievers is underlined.}\label{table:1}
\end{table*}

\section{Experimental Methodology}
\label{sec:experimental_methodology}
This section introduces the training dataset, evaluation, baseline models, and details on implementation.

\textbf{Training and Evaluation Datasets.}
We use MS MARCO~\cite{nguyen2016ms} and BEIR~\cite{thakur2021beir}, two widely used English retrieval benchmarks, to evaluate the retrieval effectiveness in both \textbf{unsupervised} and \textbf{finetuned} settings. We collect anchor document pairs from ClueWeb22~\cite{overwijk2022clueweb22}, a dataset containing extensive web page information, to train dense retrievers in an unsupervised setting. We use the training set of MS MARCO Passage as data for fine-tuning. Following previous work~\cite{thakur2021beir,xie2023unsupervised}, nDCG@10 is used as our evaluation metric.

\textbf{Implementation Details.} Our implementation is based on OpenMatch~\cite{liu2021openmatch,yu2023openmatch}. During training the \textbf{embedding model} for clustering, we think the URL of web pages contains valuable information about the relations between web pages. Therefore, we incorporate the URL with the title and content as the text representation of web pages. We cluster training data into $500$ groups, yet the actual data may contain slightly less than 500 groups because the groups smaller than $128$ are merged (as mentioned in Sec.~\ref{model:cluster}). We analyzed the effect of using different group numbers in Appendix~\ref{sec:app2}.

For unsupervised \textbf{retrieval models}, we start from the T5-base~\cite{raffel2020exploring} (with $220$M parameters) checkpoint and contrastively train the dense retrieval model on ClueWeb data with in-batch negatives and \texttt{BM25} retrieved negatives for 1 epoch. Then we reinitialize the group weight and train the model for another epoch using in-batch negatives and more challenging negatives generated with ANCE~\cite{xiong2020approximate}. To enhance the stability of \texttt{WebDRO}, we use gradient accumulation and update the group weights every $500$ training step (gradients of model parameters are not cached and updated every step).

For the finetuned setting, we tune base models with the training set of MS MARCO Passage with in-batch negatives and \texttt{BM25} retrieved negatives for 1 epoch. The finetuning process doesn't involve clustering or reweighting.

\textbf{Baseline Models.} We compare \texttt{WebDRO} with a commonly used sparse retriever, \texttt{BM25}~\cite{robertson2009probabilistic}, and five unsupervised dense retrieval models. \texttt{Anchor-DR}~\cite{xie2023unsupervised} serves as the baseline for direct comparison with \texttt{WebDRO}. \texttt{Anchor-DR} regards anchor texts as queries and uses anchor-document pairs without clustering. For the other four baselines, we describe them in Appendix~\ref{appendix:A}. 

To evaluate \texttt{WebDRO} in a fine-tuned setting, we compare three different base models: \texttt{T5-base}, \texttt{Anchor-DR}, and \texttt{WebDRO}. These base models are finetuned in exactly the same manner.

\footnotetext[1]{Average score of 14 public datasets in BEIR.}

\section{Evaluation Results}
In this section, we first show the performance of different retrieval models and analyze different clustering methods in our approach. Then we compare \texttt{WebDRO}'s performance in different tasks and analyze the validity of the group weights learned by \texttt{WebDRO}.

\subsection{Overall Performance}
Table~\ref{table:1} shows the unsupervised retrieval results of \texttt{WebDRO} on MS MARCO and BEIR. 

In the unsupervised setting, \texttt{WebDRO} outperforms \texttt{Anchor-DR}, its counterpart without reweighting by achieving a 1.2\% improvement on MS MARCO and a 1.1\% improvement on the average of BEIR. By utilizing group weights, \texttt{WebDRO} enhances retrieval performance on most datasets compared with \texttt{Anchor-DR}, indicating its effectiveness in identifying and highlighting rare or challenging data patterns in anchor-document pairs. \texttt{WebDRO} also significantly outperforms the first three unsupervised baselines. Although the amount of data used to train \texttt{WebDRO} is only 5.1\% of that used to train \texttt{E5-Large}, \texttt{WebDRO}'s average score only falls behind by 2.1\%. This indicates that anchor-document pairs are a high-quality resource for training unsupervised retrievers. We believe stronger dense retrievers can be trained in the same manner as \texttt{WebDRO} after scaling up, and we leave it for future work.

As shown in the right part of Table~\ref{table:1}, \texttt{WebDRO} also significantly outperforms \texttt{Anchor-DR} and \texttt{T5-base} on most datasets when they are all finetuned with the same MS MARCO data. Additionally, finetuned \texttt{WebDRO} outperforms \texttt{BM25} on 10 sub-tasks (out of 19) and has an average performance gain of 2.6\%. The performance gain of \texttt{WebDRO} further indicates the advantages of our clustering and reweighting method can be brought to the fine-tuned scenarios.



\begin{table}
    \centering
    \small
    \renewcommand{\arraystretch}{1.1}
    \begin{tabular}{lcc}
        \hline
        \textbf{Model} & \textbf{MS MARCO} & \textbf{Avg. on BEIR} \\\hline
        Anchor-DR & 26.15 & 39.50 \\
        \hline
        DRO w/URL  & 26.97 & 39.79 \\
        DRO w/Anchor-DR & 26.64 & 39.65 \\
        WebDRO & 27.37 & 40.60 \\
    \hline
    \end{tabular}
    \caption{Performance of variants of WebDRO using different clustering strategies in an unsupervised setting.}\label{table:ex}
\end{table}

\subsection{Analyses on Clustering Strategy}\label{analysis}
To compare the effectiveness of different document clustering strategies, we also implement two variants of \texttt{WebDRO} by using different clustering strategies. \texttt{DRO w/URL} directly clusters the web pages that share the same domain name into the same group. \texttt{DRO w/Anchor-DR} uses \texttt{Anchor-DR} to encode the documents for clustering instead of using a task-specifically trained embedding model. The results in the unsupervised setting are shown in Table~\ref{table:ex}. 

From the table, we can see \texttt{WebDRO} outperforms \texttt{DRO w/Anchor-DR} by achieving about a 1\% improvement. We hypothesize that this disparity arises from the fact that the clustering task concerns the relations between \textbf{documents}, whereas \texttt{Anchor-DR} is trained with \textbf{anchor-document pairs}, emphasizing the modeling of relations between queries and documents. A link predictor style embedding model (the one in \texttt{WebDRO}) focuses more on extracting document features to model the relation between web pages, making it more tailored for clustering documents. 


URLs are commonly employed to encapsulate the content of a web page, and naturally, \texttt{DRO w/URL} exhibits an improvement over vanilla \texttt{Anchor-DR}. However, in comparison to \texttt{DRO w/URL}, \texttt{WebDRO} shows better retrieval performance, illustrating that titles and contents also play a crucial role in enhancing the URL-based representations of web pages. Another reason for \texttt{DRO w/URL}'s performance decline compared to \texttt{WebDRO} is its coarse clustering for the documents belong to large domains. For instance, all web pages within Wikipedia will be clustered in one single group (this group contains roughly $1$M anchors, constituting $7.2\%$ of whole training data) regardless of their topics. 



\begin{figure}
  \centering
  \includegraphics[width=\linewidth]{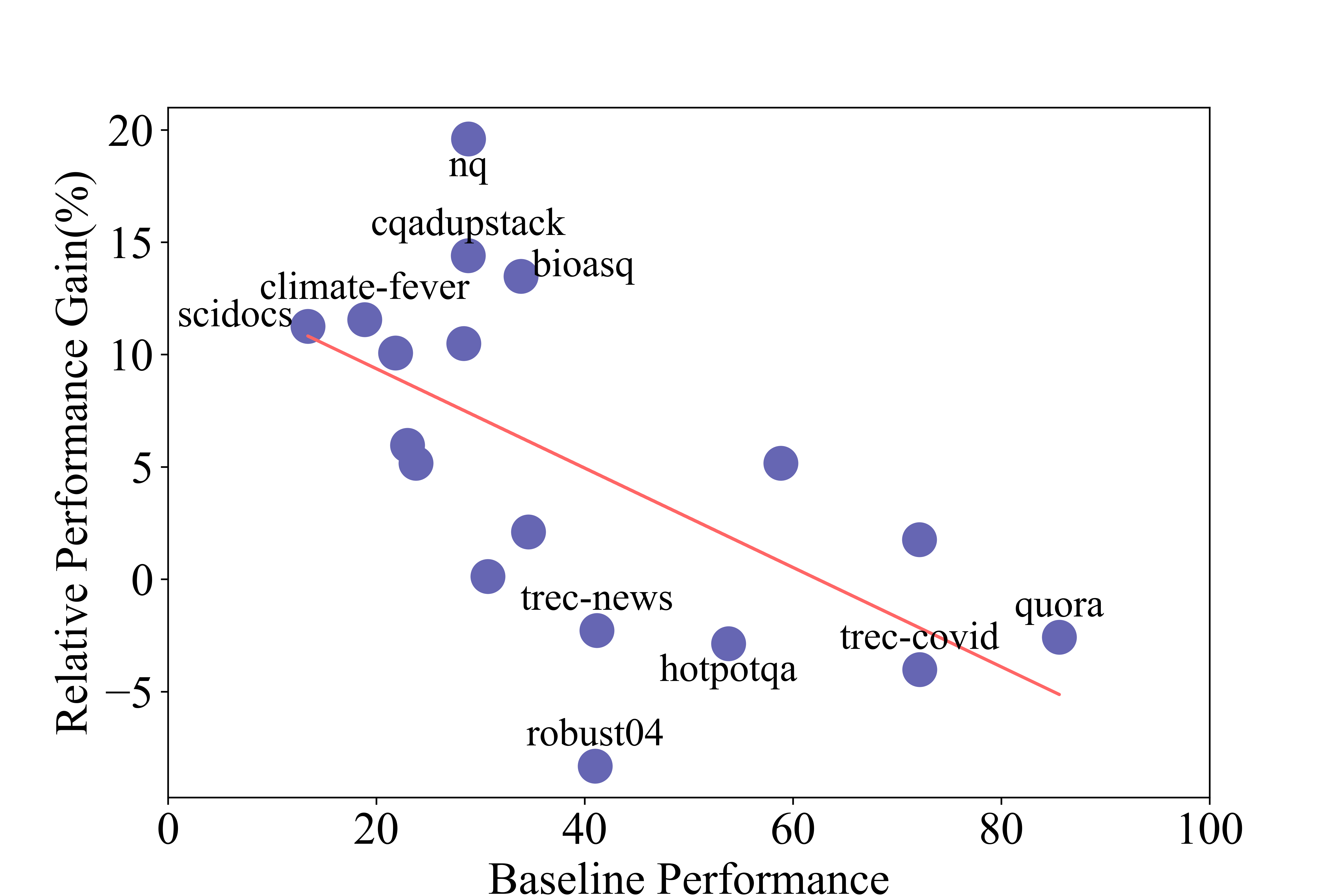}
  \caption{The relationship between performance gain of \texttt{WebDRO} and performance of \texttt{Anchor-DR}. Each point represents a dataset in BEIR. Datasets with high or negative gains are annotated.}\label{fig:5}
\end{figure}

\subsection{Retrieval Effectiveness in Different Testing Scenarios}
This section focuses on evaluating the role of the group weighting mechanism in training dense retrieval models. In Figure~\ref{fig:5}, we present the performance points of all 18 datasets from BEIR and show the relationship between \texttt{Anchor-DR}'s performance and the relative performance gain achieved by \texttt{WebDRO} over \texttt{Anchor-DR}.

The evaluation results show that the performance gap between \texttt{Anchor-DR} and \texttt{WebDRO} tends to be larger when the baseline nDCG@10 is low. It illustrates that \texttt{WebDRO} performs particularly well on difficult retrieval tasks, like \textit{scidocs} (with a relative gain of 11.26\%), \textit{hotpotQA} (19.60\%) and \textit{cqadupstack} (14.39\%). Conversely, on datasets where the baseline already exhibits strong performance, such as \textit{trec-covid} (-4.30\%) and \textit{quora} (-2.58\%), the model's enhancements are marginal or even negative. This aligns with the optimization goal of \texttt{WebDRO}, which is that the groups with larger training losses are assigned larger weights to improve the overall performance across all datasets. This phenomenon suggests that \texttt{WebDRO}'s unique optimization strategy effectively identifies and prioritizes challenging tasks during pretraining, offering a strategic advantage for performance gains, particularly in scenarios where traditional models struggle.

\subsection{Analyze on the Group Weights Learned by \texttt{WebDRO}}
\label{sec:54}
In this section, two experiments are conducted to show the stability and effectiveness of the group weights learned by \texttt{WebDRO}.

\begin{figure}
  \begin{subfigure}{0.52\columnwidth}
      \centering
      \includegraphics[width=0.9\linewidth]{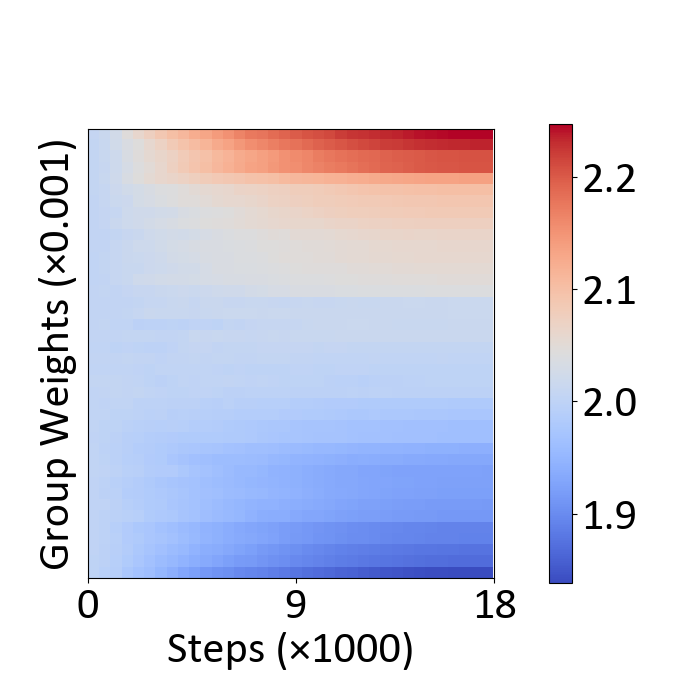}
      \caption{The weight changes of\\50 random groups.}\label{fig:2a}
  \end{subfigure}\hfill
  \begin{subfigure}{0.48\columnwidth}
      \centering
      \includegraphics[width=\linewidth]{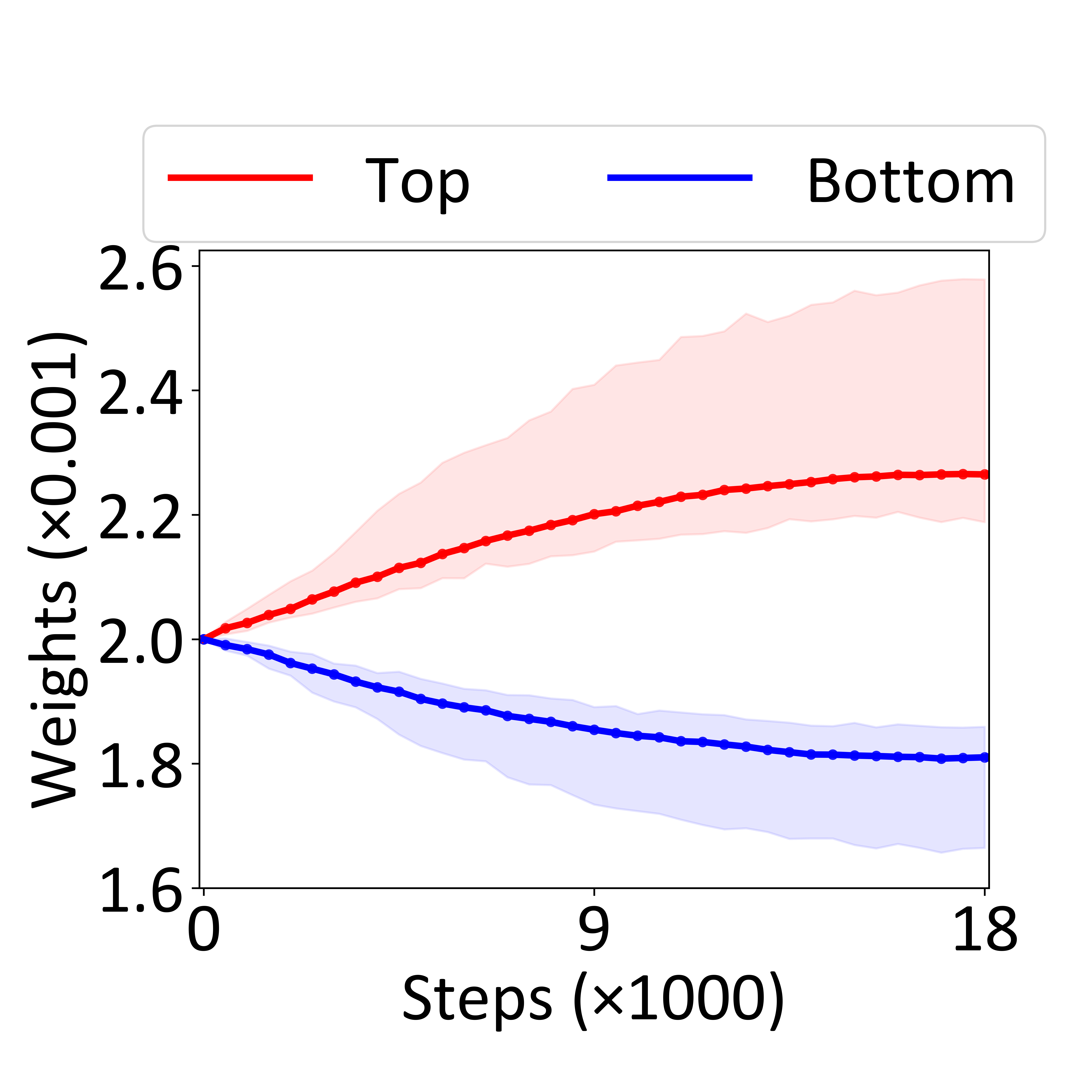}
      \caption{The weight changes of top and bottom groups.}\label{fig:2b}
  \end{subfigure}
  \caption{The visualization of group weights during training. Group weights are recorded every 500 steps.}\label{fig:2}
\end{figure}

\textbf{Trend and Distribution of Group Weights.} As shown in Figure~\ref{fig:2}, we explore the distributions of group weights that learned by \texttt{WebDRO}.

We first randomly sample 50 groups to track the changes in group weights during training. As seen in Figure~\ref{fig:2a}, the color gradient within each group gradually shifts towards darker, indicating an approximate monotonic change in group weights. Then we rank the groups according to the weights that are assigned by our final \texttt{WebDRO} model and select the top-ranked and bottom-ranked 30 groups respectively to observe the trend of weight changes during training. As shown in Figure~\ref{fig:2b}, the weights of top-weighted groups and bottom-weighted groups are consistently increased and decreased, further demonstrating the stability of the group weights learned by \texttt{WebDRO}. 

\textbf{Performance of Retrieval Models Trained with Different Weighted Groups.} We conduct experiments to pretrain dense retrieval models using groups assigned with different weights derived from \texttt{WebDRO}. Specifically, we select top and bottom 30 groups based on their weight ranks, and then sample 500k anchor-document pairs from top and bottom groups respectively. For comparison, we also train a model using a randomly sampled set of 500k pairs. Subsequently, we utilize these subsets of training data to train dense retrieval models for 1 epoch.

\begin{table}
    \centering
        \small
    \renewcommand{\arraystretch}{1.1}
    \begin{tabular}{lcc}
        \hline
        \textbf{Training Data} & \textbf{MS MARCO} & \textbf{Avg. on BEIR} \\\hline
        All Groups &  27.37 & 40.60 \\
        \hline
        Top Groups &  24.12 & 38.53 \\
        Bottom Groups  & 20.70 & 34.70 \\
        Random & 23.42 & 36.99 \\
    \hline
    \end{tabular}
     \caption{Performance of unsupervised retriever trained with different groups. Models are evaluated with nDCG@10.}\label{table:3}
\end{table}

As shown in Table~\ref{table:3}, the unsupervised retriever trained with top-weighted groups significantly outperforms other models trained with an equal number of 500k anchor-document pairs. This illustrates that the groups assigned higher weights are more informative for the model, indicating that they are underrepresented in the training data and post greater challenges. The retrieval performance of the model that is trained with randomly sampled training data significantly decreases when we replace the training data with instances in bottom-weighted groups. This highlights the capacity of \texttt{WebDRO} to effectively identify data already well-interpreted by the model.

\subsection{Analyses of the Loss Landscapes}
Loss landscape is a common tool for assessing a model's stability and generalization capabilities~\cite{li2018visualizing,hao2019visualizing}. Specifically, we apply small perturbations to the model's parameters from two random directions and visualize the losses after perturbations in a plane. The ``flatness'' of the loss landscape is defined as the size of the region around the origin where the training loss remains low~\cite{hochreiter1997flat}. A flatter loss landscape usually indicates better robustness to noises and a higher chance of converging to the optimal when randomly initialized.

\begin{figure}
  \begin{subfigure}{0.48\columnwidth}
      \centering
      \includegraphics[width=\linewidth]{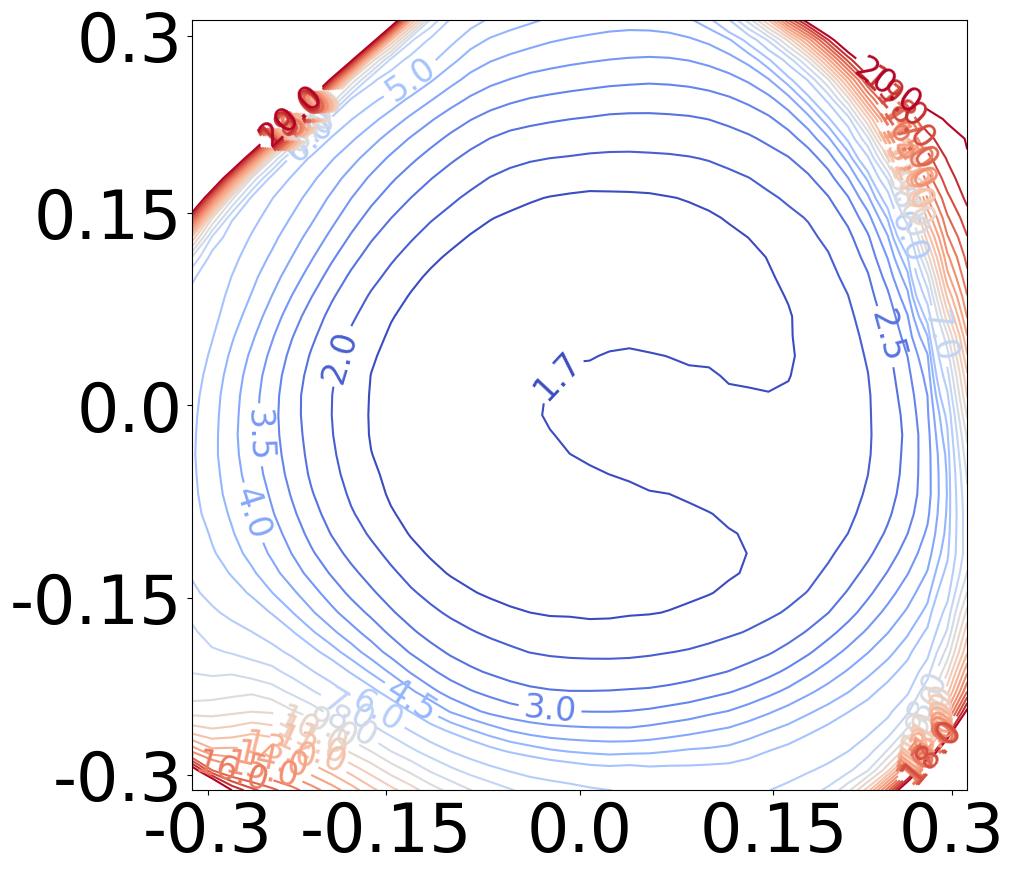}
      \caption{\texttt{Anchor-DR}.}
  \end{subfigure}\hfill
  \begin{subfigure}{0.48\columnwidth}
      \centering
      \includegraphics[width=\linewidth]{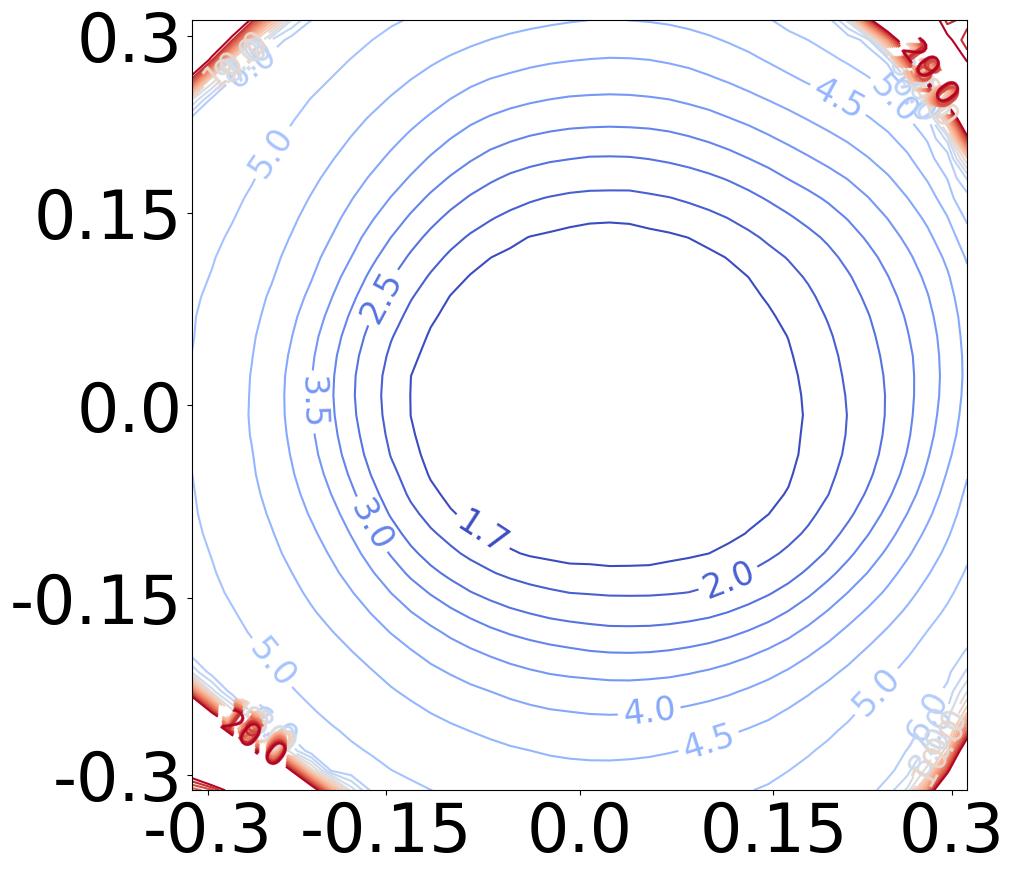}
      \caption{\texttt{WebDRO}.}
  \end{subfigure}
  \caption{Loss landscapes of \texttt{Anchor-DR} and \texttt{WebDRO}. We constrain the perturbation in each direction to $(-0.3, 0.3)$. Each closed curve in the figure indicates that the corresponding positions have the same loss.}\label{fig:6}
\end{figure}

As shown in Figure~\ref{fig:6}, we can see that with increasing perturbation, the training loss in \texttt{WebDRO} exhibits a gradual increase in all directions, whereas in \texttt{Anchor-DR}, the rise in training loss is steep in specific directions. Besides, it can be observed that \texttt{WebDRO} encompasses a larger area with low training loss (the area inside the red lines) than Anchor-DR. In summary, \texttt{WebDRO} presents a flatter and more uniform landscape than \texttt{Anchor-DR}, indicating enhanced stability in training and heightened generalization capabilities to different tasks.

\section{Conclusion}
This paper introduces \texttt{WebDRO}, an unsupervised dense retrieval model that utilizes web structures to cluster documents and adjusts group weights during contrastive training. The training process is divided into two steps. In \texttt{Step-A}, we use structures of the web graph to train a link prediction style embedding model for clustering. In \texttt{Step-B}, we utilize GroupDRO to reweight the clusters during training based on training loss. Experiments on MS MARCO and BEIR show that \texttt{WebDRO} achieves significant improvements over baseline models, especially on difficult tasks. 
Further analysis shows that group weights learned by \texttt{WebDRO} are stable and valid, which means underrepresented and challenging groups are assigned higher weights consistently throughout training. This confirms \texttt{WebDRO}'s effectiveness in improving dense retrievers' robustness.

\section*{Limitations}
An important part of \texttt{WebDRO} is training the embedding model and generating clusters. However, we only use the cluster information in reweighting, which is a waste of resources. Research has shown that introducing more challenging negatives would enhance the model's performance~\cite{xiong2020approximate}. As being in the same group indicates similarity, we hope to build stronger negatives with clusters to achieve better performance.

In this work, we cluster different contrastive pairs solely by their semantic features, which means contrastive pairs with similar content will be clustered in the same group. This approach is based on a hypothesis: semantically resembled contrastive pairs have similar contributions to retrieval model training. However, other factors may also influence the data point's contribution, like format, style, or even features we can't explicitly describe.

Data imbalance in unsupervised training data is the key motivation of our method. Although prior work has confirmed this phenomenon and its consequences, we didn't closely examine data distribution in our scenario due to our data's size.



\bibliography{custom}

\clearpage
\newpage
\appendix
\appendix
\section{Appendix}
\subsection{License}
Our code base, OpenMatch, uses MIT License. For datasets used in this work, MS MARCO can be used freely without a license, BEIR uses Apache-2.0 license and ClueWeb22 uses an individual agreement, which we have already obtained.

\subsection{Training Details}
\label{appendix:A}
This subsection describes the training details of our \texttt{WebDRO} model, including data processing, hyperparameters and computational resources.

\textbf{Data Processing.} All training data are selected from the English documents in the B category of ClueWeb22. We follow~\cite {xie2023unsupervised} to process the training data by filtering out uninformative anchors (Homepage, Login, etc.), in-domain anchors, headers, and footers using heuristic rules. A miniBERT-based classifier model is also utilized to evaluate the quality of anchor texts. We only retain 25\% anchor texts that are top-ranked. 

\textbf{Sizes of Datasets.} The training data for training unsupervised the \textbf{retrieval model} contains $13.8$M contrastive pairs, and the training data for the \textbf{embedding model} contains $5.1$M contrastive pairs. For evaluation benchmarks, MS MARCO contains $8.8$M documents and $8,960$ queries, and BEIR contains $44.0$M documents and $47,282$ queries in total.

\textbf{Hyperparameters.} The learning rate and batch size are set to $3e-5$ and $384$ for training the embedding model. During the stage of training retrieval models, we set the learning rate of the contrastive loss to $3e-5$ and batch size to $768$. The learning rate for weight updates is set to $3e-4$.

\textbf{Computational Resources.} The training of our model uses 8 NVIDIA RTX A6000 GPUs (48GB of memory each). Each epoch of dense retrieval model training (for all $13.8$M contrastive pairs) takes 9 to 12 hours depending on the inter-node communication efficiency.

\subsection{More Details of Baseline Models} \textit{coCondenser}~\cite{gao2021unsupervised} bases on the Condenser model~\cite{gao2021condenser} and is trained with token-level and span-level language modeling objectives. \textit{SPAR}~\cite{chen2021salient} further trains dense retrievers on Wikipedia and aims at imitating a sparse retrieval teacher. \textit{AugTriever}~\cite{meng2022augtriever} focuses on extracting the queries for the documents and using them to train retrieval models. \textit{E5-Large}~\cite{wang2022text} trains retrieval models on CC-Pair, a huge unsupervised dataset collected from multiple sources and curated with a consistency-based filter.

\subsection{Analyse on the Number of Clusters}
\label{sec:app2}
\begin{table}
    \centering
    \small
    \begin{tabular}{rcc}
        \hline
        \textbf{\#Group} &\textbf{MS MARCO} & \textbf{Average on BEIR} \\ 
        \hline
        50 & 26.93 & 40.40 \\
        200 & 27.43 & 40.48 \\
        500 & 27.37 & 40.70 \\
        1,000 & 27.11 & 40.49 \\
        5,000 & 26.82 & 40.17 \\
        15,000 & 27.29 & 40.00 \\
    \hline
    \end{tabular}
    \caption{Retrieval performance of \texttt{WebDRO} with different group numbers (\#Group). Results are in nDCG@10.}\label{table:2}
\end{table}
As shown in Sec.~\ref{model:cluster}, we use K-Means to cluster the documents and the number of groups ($n$) is critical for GroupDRO. We conducted a series of experiments to investigate the impact of varying $n$ values on model performance. In Table~\ref{table:2}, we can see the model performance initially is improved but later decreases as the $n$ increases. In our training scenario, $n=500$ is an appropriate value for \texttt{WebDRO} to achieve the best retrieval performance. If $n>500$, the number of the clustered groups is too large and the weights of smaller groups are usually unstable or even faulty updated. If $n<500$, the clustering is too coarse, and detailed features of documents cannot be represented through clustering. Each group contains a large number of documents, which should be assigned different weights.

\end{document}